\documentclass[12pt,a4paper,titlepage]{article}
\usepackage{fullpage}
\usepackage{amssymb,amsmath,stmaryrd}
\usepackage[mathscr]{eucal}
\usepackage{epsfig}

\newtheorem{definition}{Definition}
\newtheorem{theorem}{Theorem}
\newtheorem{proposition}{Proposition}

\newtheorem{lemma}{Lemma}

\newcommand{\Q}{\mathcal{Q}}
\def \cho {\mathcal{C}}

\newcounter{remark}
\newenvironment{remark}
{\begin{quote}\textsc{Remark} \stepcounter{remark} \arabic{remark}:}
{\end{quote}}
\newcounter{example}

\newenvironment{proof}{\medskip\noindent \bf Proof: \rm}{\hspace*{\fill}
$\blacksquare$ \newline \medskip}

\begin{document}

\title{Bi-capacities\\ Part I: Definition, M\"obius Transform and
    Interaction\thanks{Corresponding author: Michel GRABISCH, Universit\'e
    Paris I - Panth\'eon-Sorbonne.  Tel (+33) 1-44-27-88-65, Fax (+33)
    1-44-27-70-00. Email \texttt{Michel.Grabisch@lip6.fr}}}

\author{Michel GRABISCH\\
Universit\'e Paris I - Panth\'eon-Sorbonne\\
\normalsize email \texttt{Michel.Grabisch@lip6.fr}
\and
Christophe LABREUCHE\\
Thales Research \& Technology\\
Domaine de Corbeville, 91404 Orsay Cedex, France\\
\normalsize email \texttt{Christophe.Labreuche@thalesgroup.com}}

\date{}

\maketitle

\begin{abstract}
  Bi-capacities arise as a natural generalization of capacities (or fuzzy
  measures) in a context of decision making where underlying scales are
  bipolar.  They are able to capture a wide variety of decision behaviours,
  encompassing models such as Cumulative Prospect Theory (CPT). The aim of this
  paper in two parts is to present the machinery behind bi-capacities, and thus
  remains on a rather theoretical level, although some parts are firmly rooted
  in decision theory, notably cooperative game theory. The present first part
  is devoted to the introduction of bi-capacities and the structure on which
  they are defined. We define the M\"obius transform of bi-capacities, by just
  applying the well known theory of M\" obius functions as established by Rota
  to the particular case of bi-capacities. Then, we introduce derivatives of
  bi-capacities, by analogy with what was done for pseudo-Boolean functions
  (another view of capacities and set functions), and this is the key point to
  introduce the Shapley value and the interaction index for bi-capacities. This
  is done in a cooperative game theoretic perspective.  In summary, all
  familiar notions used for fuzzy measures are available in this more general
  framework.

\end{abstract}

\section{Introduction}
Capacities \cite{cho53}, also known under the name of fuzzy measures
\cite{sug74}, have become an important tool in decision making these last two
decades, allowing to model the behaviour of the decision maker in a flexible
way.  Numerous works have been done in decision under risk and uncertainty,
after the seminal work of Schmeidler \cite{sch89}, and in multicriteria
decision making (see \cite{grlava99} for a general construction based on
capacities). In the latter field, the notion of Shapley value \cite{sha53},
borrowed from cooperative game theory, and of interaction index for a pair of
criteria \cite{muso93}, have become of primary importance for the
interpretation of capacities. Later, Grabisch proposed a generalization of the
interaction index, viewing it as a linear transform on the set of capacities,
as it is also for the M\"obius transform, and permitted by this the
introduction of $k$-additive capacities, a concept which has revealed to be
very useful in applications \cite{gra96f}.

\medskip

Although being able to capture a wide variety of decision behaviours,
capacities may reveal inefficient in some situations, in particular when the
underlying scales are \emph{bipolar}. Let us introduce some formalization to go
ahead in our explanation, and choose as framework multicriteria decision
making. We consider a set $N:=\{1,\ldots,n\}$ of criteria. To simplify our
exposition we assume that to each alternative is assigned a vector of
\emph{scores} $(a_1,\ldots,a_n)$, $a_i\in[0,1]$, such that $a_i$ expresses to
which degree the alternative satisfies criterion $i$. We make the assumption
that all the scores are \emph{commensurable}, i.e., $a_i=a_j$ iff the intensity
of satisfaction for the decision maker is the same on criteria $i$ and $j$ (see
\cite{grlava99} for a complete exposition). We define a capacity $\nu$ on $N$,
i.e., a set function $\nu:2^N\longrightarrow [0,1]$ being monotone w.r.t
inclusion, and fulfilling $\nu(\emptyset)=0, \nu(N)=1$. Roughly speaking,
$\nu(A)$ expresses the degree to which the coalition of criteria $A\subseteq N$
is important for making decision.  More precisely, $\nu(A)$ is exactly the
overall score assigned to the alternative whose vector of score is
$(1_A,0_{A^c})$, i.e., all criteria in $A$ have a score equal to 1 (total
satisfaction), and all others have a score equal to 0 (no satisfaction). Such
alternatives are called \emph{binary}.  A natural way to compute the overall
score for any alternative is to use the Choquet integral $\cho_\nu$, since it
coincides with the capacity $\nu$ for binary alternatives, i.e.,
$\cho_\nu(1_A,0_{A^c})=\nu(A)$, and performs the simplest possible linear
interpolation between binary alternatives \cite{gra03b}.

However, in many practical cases, it happens that scores should be better
expressed on a \emph{bipolar scale}. Studies in psychology (see, e.g., Osgood
\emph{et al.} \cite{ossuta57}) have shown that most often scales used to
represent scores  should be considered as \emph{bipolar}, since
decision making is often guided by \emph{affect}. Quoting Slovic
\cite{slfipema02}, affect is the ``\emph{specific quality of ``goodness'' and
``badness'', as it is felt consciously or not by the decision maker, and
demarcating a positive or negative quality of stimulus}''. Then it is natural to
use a scale going from negative (bad) to positive (good) values, including a
central \emph{neutral} value, to encode the bipolarity of the affect. Such a
scale is called a \emph{bipolar scale}, typical examples are $[-1,1]$ (bounded
cardinal), $\mathbb{R}$ (unbounded cardinal) or \{very bad, bad, medium, good,
excellent\} (ordinal).

The problem is then to generalize the above construction, i.e., to define
importance of coalitions of criteria, and secondly the way of computing the
overall score of any alternative. Let us take for simplicity the $[-1,1]$
scale, with neutral value 0. The simplest way is to say that ``positive'' and
``negative'' parts are symmetric, so that the overall score of \emph{positive}
binary alternative $(1_A,0_{A^c})$ is the opposite of the one of
\emph{negative} binary alternative $(-1_A,0_{A^c})$. This leads to the
symmetric Choquet integral. A more complex model would consider only
independence between positive and negative parts, that is to say, positive
binary alternatives define a capacity $\nu_+$, while negative binary
alternatives define a different capacity $\nu_-$.  This leads to the well known
\emph{Cumulative Prospect Theory (CPT)} model, of Tversky and Kahnemann
\cite{tvka92}. Despite the generality of such models, it is not difficult to
find examples where the preference of the decision maker cannot be cast in CPT
(see \cite{lagr02,lagr03}). We can propose a yet more general model, by
considering that independence between positive and negative parts does not hold,
so that we have to consider \emph{ternary alternatives} $(1_A,-1_B,0_{(A\cup
B)^c})$, and assign to each of them a number in $[-1,1]$.  We denote this
number as $v(A,B)$, i.e., a two-argument function, whose first argument is the
set of totally satisfied criteria, and the second one the set of totally
unsatisfied criteria, the remaining criteria being at the neutral level. We
call this function \emph{bi-capacity}, since it plays the role of a capacity,
but with two arguments corresponding to the positive and negative sides of a
bipolar scale.

Interestingly enough, similar concepts have already been proposed in the field
of cooperative game theory. Bilbao \cite{bil00} has proposed
\emph{bi-cooperative games}, which coincide with our definition of bi-capacities,
although being based on a different underlying structure. \emph{Ternary
voting games} of Felsenthal and Machover \cite{fema97} are a particular case of
bi-cooperative games. Also, independently, Greco \emph{et al.} have proposed
\emph{bipolar capacities} \cite{grmasl02}, where they consider that $v(A,B)$ is
a pair of real numbers (we will address them in the second part of our paper).

\medskip

Our aim in this two-parts paper is to settle down the machinery of
bi-capacities, so that it can serve as a departure for a new area in decision
making and game theory. Hence we will remain on an abstract level, trying to
find equivalent notions to what is already known and useful for capacities and
cooperative games. In the first part of this paper, our aim is to study the
structure on which bi-capacities are defined (Section 4), to introduce the
M\"obius transform of bi-capacities as well as $k$-additive bi-capacities
(Section 5), and the derivative of bi-capacities (Section 6). We turn then to
bi-cooperative games, which are more general since no monotonicity is assumed,
and we define the Shapley value and the interaction index (Section 7). The second
part of the paper will be essentially devoted to the definition of the Choquet
and Sugeno integrals.

Throughout the paper, $N:=\{1,\ldots,n\}$ denotes the finite referential set.
To avoid heavy notations, we will often omit braces and commas to denote
sets. For example, $\{i\}, \{i,j\}, \{1,2,3\}$ are respectively denoted by $i,
ij, 123$. Cardinality of sets will be often denoted by the corresponding lower
case, e.g., $n$ for $|N|$, $k$ for $|K|$, etc.

\section{Preliminaries}
\label{sec:prel}
We begin by recalling basic notion about capacities \cite{cho53} (also called
\emph{fuzzy measures} by Sugeno \cite{sug74}) for finite sets. 

A (cooperative) \emph{game} $\nu:2^N\longrightarrow \mathbb{R}$ is a set
function such that $\nu(\emptyset)=0$. A \emph{capacity} $\nu$ is a game such
$A\subseteq B\subseteq N$ implies $\nu(A)\leq\nu(B)$. The capacity is
\emph{normalized} if in addition $\nu(N)=1$. The \emph{conjugate capacity} of a
normalized capacity $\nu$ is the normalized capacity $\bar{\nu}$ defined by
$\bar{\nu}(A):=1-\nu(N\setminus A)$ for every $A\subseteq N$. A capacity $\nu$
is additive if $\nu(A) = \sum_{i\in A}\nu(\{i\}) $, for every $A\subseteq N$.

\emph{Unanimity games} are particular capacities, defined for all $B\subseteq N$
by 
\[
u_B(A) = \left\{        \begin{array}{ll}
                        1, & \text{ if } A\supseteq B,\\
                        0, & \text{ otherwise.}
                        \end{array}     \right.
\]
Note that $u_\emptyset$ is not a capacity since $u_\emptyset(\emptyset)=1$. 

\medskip

Capacities can be viewed as special cases of \emph{pseudo-Boolean functions},
which are functions $f:\{0,1\}^n\longrightarrow\mathbb{R}$.  By making the
usual bijection between $\{0,1\}^n$ and $\mathcal{P}(N)$, any pseudo-Boolean
function $f$ on $\{0,1\}^n$ corresponds to a real-valued set function $\nu$ on
$N$ and vice-versa, with $f(1_S) \equiv \nu(S)$, $\forall S\subseteq N$, where
$1_S$ is the vector $(x_1,\ldots,x_n)\in\{0,1\}^n$, with $x_i=1$ iff $i\in
S$. Thus, capacities are non negative monotonic pseudo-Boolean functions.

\emph{Derivatives} of pseudo-Boolean functions are defined recursively as
follows. For any $\emptyset\neq S\subseteq N$, the \emph{$S$-derivative} of $f$
at point $x$ is defined by:
\begin{equation}
\Delta_Sf(x) := \Delta_i (\Delta_{S\setminus i}f(x))
\end{equation}
for any $i\in S$, with $\Delta_if(x) :=
f(x_1,\ldots,x_{i-1},1,x_{i+1},\ldots,x_n) -
f(x_1,\ldots,x_{i-1},0,x_{i+1},\ldots,x_n)$, and $\Delta_{\emptyset}f =
f$. This definition is unambiguous, and $\Delta_Sf$ depends no more on the
variables contained in $S$. Hence, one can speak of the \emph{derivative of a
capacity $\nu$} w.r.t.         subset $S$ at point $T$. The explicit formula is:
\begin{equation}
\Delta_S \nu(T) = \sum_{L\subseteq S}(-1)^{s-l}\nu(L\cup T), \forall
S\subseteq N,\forall T\subseteq N\setminus S.
\end{equation}

\medskip

As lattices are of central concern in this paper, we briefly recall elementary
definitions and useful results (see, e.g., \cite{dapr90} for details). A set $L$
endowed with a reflexive, antisymmetric and transitive relation $\leq$ is a
\emph{lattice} if for every $x,y\in L$, a unique least upper bound (denoted
$x\vee y$) and a unique greatest lower bound $x\wedge y$ exist. The \emph{top}
$\top$ (resp. \emph{bottom} $\bot$) of $L$ is the greatest (resp. the least)
element of $L$, and always exists when the lattice is finite. A lattice is
\emph{distributive} when $\vee,\wedge$ satisfy the distributivity law, and it
is \emph{complemented} when each $x\in L$ has a (unique) complement $x'$,
i.e., satisfying $x\vee x'=\top$ and $x\wedge x'=\bot$. A lattice is said to be
\emph{Boolean} if it has a top and bottom element, is distributive and
complemented. When $L$ is finite, it is Boolean iff it is isomorphic to the
lattice $2^n$ for some $n$.

$Q\subseteq L$ is a \emph{down-set} of $L$ if $x\in Q$ and $y\leq x$ implies
$y\in Q$. For any $x\in L$, the \emph{principal ideal} $\downarrow x$ is
defined as $\downarrow x:=\{y\in L\mid y\leq x\}$ (down-set generated by
$x$). More generally, for $A\subseteq L$, $\downarrow A:=\bigcup_{x\in
A}\downarrow x$. Similar definitions exist for \emph{up-sets} and
\emph{principal filters} $\uparrow x$. For $x,y\in L$, we say that $x$
\emph{covers} $y$ (or $y$ is a \emph{predecessor} of $x$), denoted by $x\succ y$,
if there is no $z\in L, z\neq x,y$ such that $x\leq z\leq y$.  An element $i\in
L$ is \emph{join-irreducible} if it cannot be written as a supremum over other
elements of $L$. When $L$ is finite, this is equivalent to $i$ covers only one
element. \emph{Atoms} are join-irreducible elements covering $\bot$. We call
$\mathcal{J}(L)$ the set of all join-irreducible elements of $L$.

In a finite distributive lattice, any element $y\in L$ can be decomposed in
terms of join-irreducible elements. The fundamental result due to Birkhoff is
the following \cite{bir33}.
\begin{theorem}
\label{th:birk}
Let $L$ be a finite distributive lattice. Then the map $\eta:L\longrightarrow
\mathcal{O}(\mathcal{J}(L))$, where $\mathcal{O}(\mathcal{J})$ is the set of all
down-sets of $\mathcal{J}$, defined by
\[
\eta(x) :=\{i\in\mathcal{J}(L)\mid i\leq x\}=\mathcal{J}(L)\cap \downarrow x
\]
is an isomorphism of $L$ onto $\mathcal{O}(\mathcal{J}(L))$.
\end{theorem}
We call $\eta(x)$ the \emph{normal decomposition} of $x$, we have 
\[
x=\bigvee \eta(x).
\] 
The decomposition of some $x$ in $L$ in terms of a supremum of join-irreducible
elements is unique up to the fact that it may happen that some join-irreducible
elements in $\eta(x)$ are comparable. Hence, if $i\leq j$ and $j$ is in a
decomposition of $x$, then we may delete $i$ in the decomposition. We call
\emph{irredundant decomposition} the (unique) decomposition of minimal
cardinality, and denote it by $\eta^*(x)$. It is unique whenever the lattice is
distributive.

\section{Bi-capacities}
Let us denote  $\Q(N):=\{(A,B)\in
\mathcal{P}(N)\times\mathcal{P}(N) |A\cap B=\emptyset\}$, where
$\mathcal{P}(N)$ stands for $2^N$.
\begin{definition}
\label{def:bica}
A function $v:\mathcal{Q}(N)\longrightarrow \mathbb{R}$ is
a \emph{bi-capacity} if it satisfies:
\begin{itemize}
\item [(i)] $v(\emptyset, \emptyset) = 0$
\item [(ii)] $A\subseteq B$ implies $v(A,\cdot)\leq v(B,\cdot)$ and
$v(\cdot,A)\geq v(\cdot,B)$. 
\end{itemize}
In addition, $v$ is \emph{normalized} if $v(N,\emptyset)=1=-v(\emptyset,N)$. 
\end{definition}
In the sequel, unless otherwise specified, we will consider that bi-capacities
are normalized. Note that the definition implies that $v(\cdot,\emptyset)\geq
0$ and $v(\emptyset,\cdot) \leq 0$.

An interesting particular case is when left and right part can be separated. We
say that a bi-capacity is of the \emph{CPT type} (refering to Cumulative
Prospect Theory \cite{tvka92}, see Introduction) if there exist two
(normalized) capacities $\nu_1,\nu_2$ such that
\[
v(A,B) = \nu_1(A) - \nu_2(B), \forall (A,B)\in \Q(N). 
\]
When $\nu_1=\nu_2$, we say that the bi-capacity is \emph{symmetric}, and
\emph{asymmetric} when $\nu_2 = \overline{\nu}_1$. 

By analogy with the classical case, a bi-capacity is said to be
\emph{additive} if it is of the CPT type with $\nu_1,\nu_2$ being additive,
i.e., it satisfies for all $(A,B)\in \mathcal{Q}(N)$:
\begin{equation}
\label{eq:1}
v(A,B) = \sum_{i\in A}\nu_1(\{i\}) - \sum_{i\in B}\nu_2(\{i\}).
\end{equation}
Since for an additive capacity, $\overline{\nu}=\nu$, an additive bi-capacity
with $\nu_1=\nu_2$ is both symmetric and asymmetric.

 More generally, decomposable bi-capacities can be defined as well, using
t-conorms (see \cite{grla02b}) or uninorms with neutral element 0 (see
\cite{same03}), we do not develop this topic here.

\section{The structure of $\Q(N)$}
\label{sec:struc}
We study in this section the structure of $\Q(N)$. From its definition, $\Q(N)$
is isomorphic to the set of mappings from $N$ to $\{-1,0,1\}$, hence
$|\Q(N)|=3^n$. Also, any element $(A,B)$ in $\Q(N)$ can be denoted by
$(x_1,\ldots,x_n)$, with $x_i\in \{-1,0,1\}$, and $x_i=1$ if $i\in A$,
$x_i=-1$ if $i\in B$, and 0 otherwise. 
 
As a preliminary remark, $\Q(N)$ is a subset of $\mathcal{P}(N)^2$, and can
therefore be represented in a matrix form, using some total order on
$\mathcal{P}(N)$. A natural order is the binary order, already used in
\cite{grmaro99a}, obtained by ordering in an increasing sequence the integers
coding the elements of $\mathcal{P}(N)$: $\emptyset,\{1\},\{2\},\{1,2\},\{3\},
\{1,3\}$, etc. Using this order, the matrix has a fractal structure with
generating pattern
\[
\begin{matrix}
\times & \times \\
\times &
\end{matrix}
\]
We give below the matrix obtained with $n=3$.
\[
\begin{array}{cc}
& \scriptsize{\emptyset\;\;\;\;1\;\;\;\;2\;\;\;\;12\;\;\;3\;\;\;13\;\;23\;\;123}\\
\begin{matrix} \scriptsize{\emptyset} \\ \scriptsize{1} \\ \scriptsize{2}\\
\scriptsize{12} \\ \scriptsize{3} \\ \scriptsize{13} \\ \scriptsize{23} \\
\scriptsize{123}
\end{matrix}
& \begin{bmatrix}
\times & \times & \times & \times & \times & \times & \times & \times \\
\times &        & \times &        & \times &        & \times & \\
\times & \times &        &        & \times & \times &  &\\
\times &        &        &        & \times & & & \\
\times & \times & \times & \times & & & & \\
\times &        & \times & & & & & \\
\times & \times & & & & & & \\
\times & & & & & & & 
\end{bmatrix}
\end{array}
\]
As for $\mathcal{P}(N)$, it is convenient to define a total order on $\Q(N)$,
so as to reveal structures. A natural one is to use a ternary coding. Several
are possible, but it seems that the most suitable one is to code (denoting
elements of $\Q(N)$ as $(x_1,\ldots,x_n)$, with $x_i$ in $\{-1,0,1\}$) -1 by 0,
0 by 1, and 1 by 2. The increasing sequence of integers in ternary code is 0,
1, 2, 10, 11, 12, 20 etc., which leads to the following order of elements of
$\Q(N)$:
\[
\cdots(2,3)\;(12,3)\;\boxed{(\emptyset,12)\;(\emptyset,2)\;(1,2)\;\boxed{(\emptyset,1)\;\boxed{(\emptyset,\emptyset)}\;(1,\emptyset)}\;(2,1)\;(2,\emptyset)\;(12,\emptyset)}\;(3,12)\;(3,2)\cdots
\]
Again, we remark a fractal structure, which is enhanced by boxes: the $(k+1)$th
box is built from the $k$th box by adding to its elements (of $\Q(N)$) element
$k$ of $N$, either to their left part, or to their right part.

\medskip

\begin{figure}[htb]
\begin{center}
\setlength{\unitlength}{0.7cm}
\begin{picture}(11,13)
\newsavebox{\sublatticeb}
\savebox{\sublatticeb}(5,6)[bl]{
        \put(0,2.1){\line(0,1){1.8}}
        \put(0.08,3.945){\line(3,-2){2.83}}
        \put(0.08,1.945){\line(3,-2){2.83}}
        \put(3,0.1){\line(0,1){1.8}}
        \put(3.07,0.07){\line(1,1){1.86}}
        \put(3.07,2.07){\line(1,1){1.86}}
        \put(5,2.1){\line(0,1){1.8}}
        \put(0.07,2.07){\line(1,1){1.86}}
        \put(0.07,4.07){\line(1,1){1.86}}
        \put(2,4.1){\line(0,1){1.8}}
        \put(2.08,3.945){\line(3,-2){2.83}}
        \put(2.08,5.945){\line(3,-2){2.83}}
        \put(0,2){\circle{0.2}}
        \put(0,4){\circle{0.2}}
        \put(3,0){\circle{0.2}}
        \put(3,2){\circle{0.2}}
        \put(5,2){\circle{0.2}}
        \put(5,4){\circle{0.2}}
        \put(2,4){\circle{0.2}}
        \put(2,6){\circle{0.2}}
}
\put(0,2){\usebox{\sublatticeb}}
\put(3,0){\usebox{\sublatticeb}}
\put(0,4){\usebox{\sublatticeb}}
\put(3,2){\usebox{\sublatticeb}}
\thicklines
\put(2,4){\usebox{\sublatticeb}}
\thinlines
\put(5,2){\usebox{\sublatticeb}}
\put(2,6){\usebox{\sublatticeb}}
\put(5,4){\usebox{\sublatticeb}}
\put(-0.9,8){\tiny $12,3$}
\put(-0.8,6){\tiny $1,3$}
\put(-0.9,4){\tiny $1,23$}
\put(1.1,10){\tiny $12,\emptyset$}
\put(1.3,8){\tiny $1,\emptyset$}
\put(1.3,6){\tiny $1,2$}
\put(2.9,12){\tiny $123,\emptyset$}
\put(3.1,10){\tiny $13,\emptyset$}
\put(3.1,8){\tiny $13,2$}
\put(2.3,5.9){\tiny $2,3$}
\put(2.3,3.9){\tiny $\emptyset,3$}
\put(2.1,1.9){\tiny $\emptyset,23$}
\put(5.3,8){\tiny $2,\emptyset$}
\put(5.3,6){\tiny $\emptyset,\emptyset$}
\put(5.3,3.9){\tiny $\emptyset,2$}
\put(7.2,10){\tiny $23,\emptyset$}
\put(7.2,8){\tiny $3,\emptyset$}
\put(7.2,6){\tiny $3,2$}
\put(6.3,4){\tiny $2,13$}
\put(6.3,2){\tiny $\emptyset,13$}
\put(6.3,0){\tiny $\emptyset,123$}
\put(8.3,6){\tiny $2,1$}
\put(8.3,4){\tiny $\emptyset,1$}
\put(8.3,2){\tiny $\emptyset,12$}
\put(10.3,8){\tiny $23,1$}
\put(10.3,6){\tiny $3,1$}
\put(10.3,4){\tiny $3,12$}
        \put(0,4){\circle*{0.2}}
        \put(3,2){\circle*{0.2}}
        \put(6,2){\circle*{0.2}}
        \put(6,4){\circle*{0.2}}
        \put(8,2){\circle*{0.2}}
        \put(10,4){\circle*{0.2}}
\end{picture}
\caption{The lattice $\Q(N)$ for $n=3$}
\label{fig:lat3}
\end{center}
\end{figure}
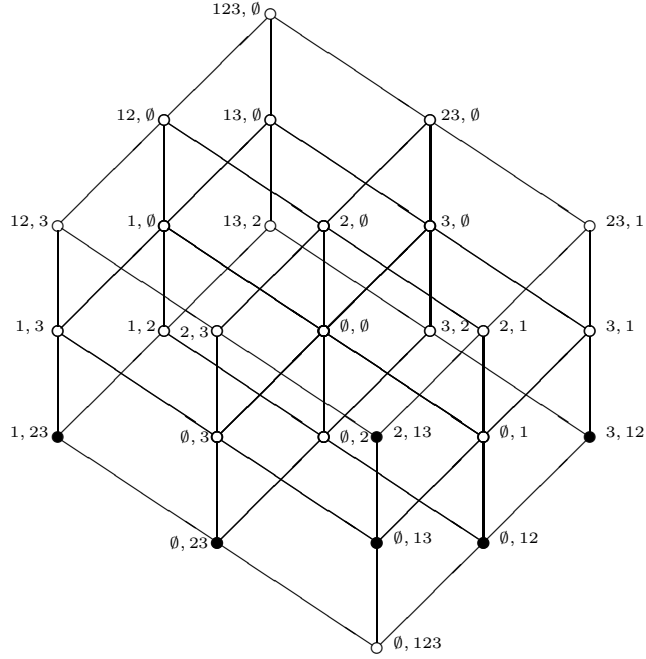

It is easy to see that $\Q(N)$ is a lattice, when equipped with the following
order: $(A,B)\sqsubseteq(C,D)$ if $A\subseteq C$ and $B\supseteq D$. Supremum and infimum
are respectively
\begin{align*}
(A,B)\sqcup(C,D)  & = (A\cup C, B\cap D)\\
(A,B)\sqcap(C,D) &  = (A\cap C, B\cup D).
\end{align*}
These are elements of $\Q(N)$ since $(A\cup C)\cap( B\cap D)=\emptyset$ and
$(A\cap C)\cap(B\cup D)=\emptyset$. Top and bottom are respectively
$(N,\emptyset)$ and $(\emptyset, N)$. Notice that a bi-capacity is an
order-preserving mapping from $\Q(N)$ to $\mathbb{R}$. We call \emph{vertex}
of $\Q(N)$ any element $(A,B)$ such that $A\cup B=N$,  they correspond to
the ``geometrical'' vertices. We give in Figure \ref{fig:lat3}
the Hasse diagram of $(\Q(N),\sqsubseteq)$ for $n=3$.

In \cite{bil00}, Bilbao \emph{et al.} introduced other operations on
$\Q(N)$, which are:
\begin{align*}
(A,B)\sqcup'(C,D) &:=((A\cup C)\setminus(B\cup D), (B\cup
D)\setminus(A\cup C))\\ 
(A,B)\sqcap'(C,D) &:=(A\cap C,B\cap D).
\end{align*}
However, $(\Q(N),\sqcup',\sqcap')$ is not a lattice since
for any $A\subseteq N, $$(A,A^c)\sqcup'(A^c,A)=(\emptyset,\emptyset)$ but
$(A,A^c)\not\sqsubseteq'(\emptyset,\emptyset)$ since
$(A,A^C)\sqcup(\emptyset,\emptyset)=(A,A^c)\neq(\emptyset,\emptyset)$. 

Following usual conventions, $\Q(N)$ is the lattice called $3^n$ (see,
e.g., \cite{dapr90}). It is formed by $2^n$ Boolean sub-lattices $2^n$: each
sub-lattice corresponds to a given partition of $N$ into two parts, one for
positive scores, the other for negative ones, which contain all subsets of
non-zero scores, including the empty set. Hence, all these sub-lattices have as
a common point $(\emptyset,\emptyset)$.

For any ordered pair $((A,B),(A\cup D, B\setminus C))$ of $\Q(N)$ with
$C\subseteq B$ and $D\subseteq (N\setminus(A\cup B))\cup C$, the interval
$[(A,B),(A\cup D, B\setminus C)]$ is a sub-lattice of type $2^k\times 3^l$,
with $k=|C\Delta D|$, and $l=|C\cap D|$. As a particular case, a sub-lattice of
type $2^k$ is obtained if $C\cap D=\emptyset$, and of type $3^l$ if $C=D$. 

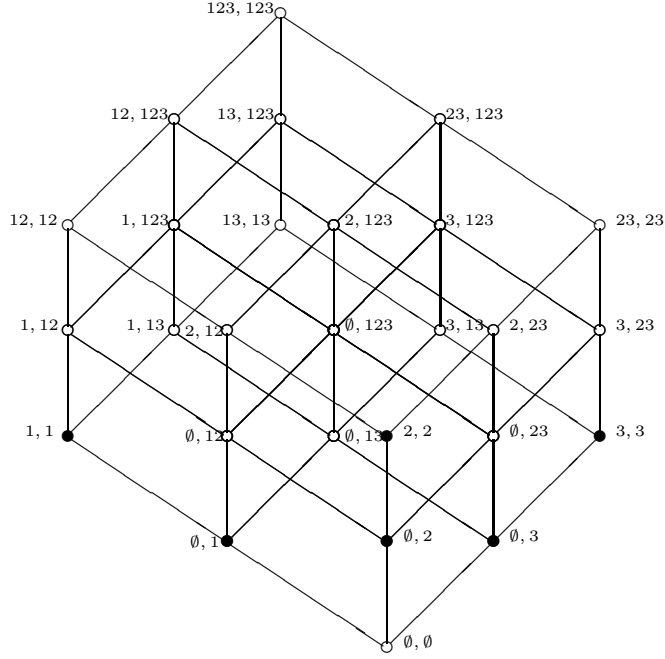
\begin{figure}[htb]
\begin{center}
\setlength{\unitlength}{0.7cm}
\begin{picture}(11,13)
\newsavebox{\sublatticebs}
\savebox{\sublatticebs}(5,6)[bl]{
        \put(0,2.07){\line(0,1){1.86}}
        \put(0.07,3.93){\line(3,-2){2.86}}
        \put(0.07,1.93){\line(3,-2){2.86}}
        \put(3,0.07){\line(0,1){1.86}}
        \put(3.07,0.07){\line(1,1){1.86}}
        \put(3.07,2.07){\line(1,1){1.86}}
        \put(5,2.07){\line(0,1){1.86}}
        \put(0.07,2.07){\line(1,1){1.86}}
        \put(0.07,4.07){\line(1,1){1.86}}
        \put(2,4.07){\line(0,1){1.86}}
        \put(2.07,3.93){\line(3,-2){2.86}}
        \put(2.07,5.93){\line(3,-2){2.86}}
        \put(0,2){\circle{0.2}}
        \put(0,4){\circle{0.2}}
        \put(3,0){\circle{0.2}}
        \put(3,2){\circle{0.2}}
        \put(5,2){\circle{0.2}}
        \put(5,4){\circle{0.2}}
        \put(2,4){\circle{0.2}}
        \put(2,6){\circle{0.2}}
}
\put(0,2){\usebox{\sublatticebs}}
\put(3,0){\usebox{\sublatticebs}}
\put(0,4){\usebox{\sublatticebs}}
\put(3,2){\usebox{\sublatticebs}}
\thicklines
\put(2,4){\usebox{\sublatticebs}}
\thinlines
\put(5,2){\usebox{\sublatticebs}}
\put(2,6){\usebox{\sublatticebs}}
\put(5,4){\usebox{\sublatticebs}}
\put(-1.1,8){\tiny $12,12$}
\put(-0.9,6){\tiny $1,12$}
\put(-0.8,4){\tiny $1,1$}
\put(0.8,10){\tiny $12,123$}
\put(1.0,8){\tiny $1,123$}
\put(1.1,6){\tiny $1,13$}
\put(2.6,12){\tiny $123,123$}
\put(2.8,10){\tiny $13,123$}
\put(2.9,8){\tiny $13,13$}
\put(2.2,5.9){\tiny $2,12$}
\put(2.2,3.9){\tiny $\emptyset,12$}
\put(2.3,1.9){\tiny $\emptyset,1$}
\put(5.2,8){\tiny $2,123$}
\put(5.2,6){\tiny $\emptyset,123$}
\put(5.2,3.9){\tiny $\emptyset,13$}
\put(7.1,10){\tiny $23,123$}
\put(7.1,8){\tiny $3,123$}
\put(7.1,6){\tiny $3,13$}
\put(6.3,4){\tiny $2,2$}
\put(6.3,2){\tiny $\emptyset,2$}
\put(6.3,0){\tiny $\emptyset,\emptyset$}
\put(8.3,6){\tiny $2,23$}
\put(8.3,4){\tiny $\emptyset,23$}
\put(8.3,2){\tiny $\emptyset,3$}
\put(10.3,8){\tiny $23,23$}
\put(10.3,6){\tiny $3,23$}
\put(10.3,4){\tiny $3,3$}
        \put(0,4){\circle*{0.2}}
        \put(3,2){\circle*{0.2}}
        \put(6,2){\circle*{0.2}}
        \put(6,4){\circle*{0.2}}
        \put(8,2){\circle*{0.2}}
        \put(10,4){\circle*{0.2}}
\end{picture}
\caption{The lattice $\Q^*(N)$ for $n=3$}
\label{fig:lat3star}
\end{center}
\end{figure}

Let us remark that the elements of $\Q(N)$ appear in a rather unnatural way on
Fig~\ref{fig:lat3}. It is possible to have a more natural structure if we
replace each element $(A,B)$ by $(A,B^c)$. Let us call this new lattice
$(\Q^*(N),\sqsubseteq^*)$. An element $(A,B)$ in $\Q^*(N)$ is
such that $A\subseteq B$, and $A$ is the set of scores equal to 1, while $B$ is
the set of scores being equal to 0 or 1. We have
\begin{align*}
(A,B)\sqsubseteq^*(C,D) & \text{ if and only if } A\subseteq C \text{ and } B\subseteq D\\
(A,B)\sqcup^* (C,D)  & = (A\cup C, B\cup D)\\
(A,B)\sqcap^* (C,D)  & = (A\cap C, B\cap D).
\end{align*}
Hence, $\sqsubseteq^*$ is simply the product order on $\mathcal{P}(N)^2$. Figure \ref{fig:lat3star} shows the Hasse diagram of
$(\Q^*(N),\sqsubseteq^*)$ for $n=3$. 

Remark also that a third alternative, we could denote by $\Q^{**}(N)$, would be
to replace in $\Q(N)$ each $(A,B)$ by $(A, (A\cup B)^c)$, the right argument
being the set of scores being equal to 0. The order relation becomes
$(A,B)\sqsubseteq^{**}(C,D)$ iff $A\subseteq C$ and $B\subseteq C\cup D$.
Although it may be mathematically more appealing to use either $\Q^*(N)$ or
$\Q^{**}(N)$, we stick in this paper to the first introduced notation, since it
is more intuitive for our original motivation of multicriteria decision making
and game theory.

Let us give some properties of $\Q(N)$ (they are the same for $\Q^*(N)$). Since
$3^n$ is a product of distributive lattices, it is itself distributive (see,
e.g., \cite{dapr90}). However it is not complemented, since for example
$(\emptyset,\emptyset)$ has no complement ($b$ is the complement of $a$ if
$a\wedge b = \bot$ and $a\vee b=\top$). It is possible to give a simpler
representation of $\Q(N)$, using join-irreducible elements (see Section
\ref{sec:prel}).  It is easy to see that the join-irreducible elements of
$\Q(N)$ are $(\emptyset,i^c)$ and $(i,i^c)$, for all $i\in N$. Since $\Q(N)$ is
distributive, the representation theorem applies, and we have for any $(A,B)\in
\Q(N)$,
\begin{equation}
\label{eq:irre}
(A,B) = \bigsqcup_{i\in A}(i,i^c)\sqcup\bigsqcup_{j\in N\setminus
B}(\emptyset,j^c) = \bigsqcup_{i\in A}(i,i^c)\sqcup\bigsqcup_{j\in N\setminus
(A\cup B)}(\emptyset,j^c).
\end{equation}
The first equality gives the normal decomposition $\eta(A,B)$, while the second
one gives the irredundant decomposition.

In $(\Q^*(N),\sqsubseteq^*)$, the join-irreducible elements are $(\emptyset,i)$
and $(i,i)$, $\forall i\in N$, while in $(\Q^{**}(N),\sqsubseteq^{**})$ they
are $(\emptyset,i)$ and $(i,\emptyset)$. On Figures \ref{fig:lat3} and
\ref{fig:lat3star}, join-irreducible elements are indicated by black circles.

Join-irreducible elements permit to define \emph{layers} in $\Q(N)$ as follows:
$(\emptyset,N)$ is the bottom layer (layer 0), the set of all join-irreducible
elements forms layer 1, and layer $k$, for $k=2,\dots,n$, contains all elements
whose irredundant decomposition contains exactly $k$ join-irreducible
elements. Layer $k$ is denoted by $\Q^{[k]}(N)$, and contains all elements
$(A,B)$ such that $|B|=n-k$, for $k=0,\ldots,n$.

\section{M\"obius transform of bi-capacities}
\label{sec:mob}
Let us recall some basic facts about the M\"obius transform (see
\cite{rot64}). Let us consider $f,g$ two real-valued functions on a locally
finite poset $(X,\leq)$ such that
\begin{equation}
\label{eq:mob1}
g(x) = \sum_{y\leq x} f(y).
\end{equation}
The solution of this equation in term of $g$ is given through the M\"obius
function $\mu$ by
\begin{equation}
\label{eq:mob2}
f(x) = \sum_{y\leq x}\mu(y,x)g(y)
\end{equation}
where $\mu$ is defined inductively by
\[
\mu(x,y) = \left\{      \begin{array}{ll}
                        1, & \text{ if } x=y\\
                        -\sum_{x\leq t< y}\mu(x,t), & \text{ if } x< y\\
                        0, &  \text{ otherwise}.
                        \end{array}     \right.
\]
Note that $\mu$ depends only on the structure of $(X,\leq)$.  When $(X,\leq)$
is a Boolean lattice, as for example $(\mathcal{P}(N),\subseteq)$, it is well
known that the M\"obius function becomes, for any $A,B\in\mathcal{P}(N)$
\begin{equation}
\mu(A,B) = \left\{      \begin{array}{ll}
                        (-1)^{|B\setminus A|} & \text{ if } A\subseteq B\\
                        0, &  \text{ otherwise}.
                        \end{array}     \right.
\end{equation}
Observe that this M\"obius function has the following property
\begin{equation}
\sum_{A\subseteq C\subseteq B}\mu(A,C) = 0, \quad \forall A,B\subseteq N, A\neq B.
\label{eq:sum}
\end{equation} 
Indeed, when $A\subsetneq B$
\begin{align*}
\sum_{A\subseteq C\subseteq B}\mu(A,C) & = \sum_{A\subseteq C\subseteq
B}(-1)^{|C\setminus A|}\\
        & = \sum_{k=0}^{|B\setminus A|} \binom{|B\setminus A|}{k}(-1)^k\\
        & = (1-1)^{|B\setminus A|} = 0.
\end{align*}

If $g$ is a capacity, which we denote by $\nu$, then $f$ in Eq. (\ref{eq:mob1})
is called the \emph{M\"obius transform} of $\nu$, usually denoted by $m$ or
$m^{\nu}$ if necessary. Equations (\ref{eq:mob1}) and (\ref{eq:mob2}) become
\begin{align}
\nu(A) = &  \sum_{B\subseteq A}m(B) \label{eq:mob3}\\
m(A) = & \sum_{B\subseteq A}(-1)^{|A\setminus B|} \nu(B).
\end{align}
Note that $m(\emptyset) = 0$.  The M\"obius transform is an important concept
for capacities and games, as it can be viewed as the coordinates of
$\nu$ in the basis of unanimity games. Indeed, Eq. (\ref{eq:mob3}) can be
rewritten as
\[
\nu(A) = \sum_{B\subseteq N} m(B) u_B(A).
\]
Note that there is a close relation with the derivative of $\nu$ since we have
\begin{equation}
\label{eq:md}
m^{\nu}(S)  =  \Delta_S \nu(\emptyset).
\end{equation}

It is a well-known result that a capacity is additive if and only if its
M\"obius transform is non zero only for singletons.  An extension of this fact
leads to the introduction of $k$-additive capacities \cite{gra96f,gra98c}. A
capacity $\nu$ is said to be \emph{$k$-additive}, for some $k$ in
$\{1,\ldots,n-1\}$, if its M\"obius transform vanishes for subsets of more than
$k$ elements, i.e., $\forall A\subseteq N$, $|A|>k$, $m(A)=0$, and there is at
least one subset $A$ such that $|A|=k$ and $m(A)\neq 0$. Clearly, 1-additive
capacities coincide with additive capacities.

We turn now to bi-capacities. The first step is to
obtain the M\"obius function on $\Q(N)$. 
\begin{theorem}
\label{th:mob}
The M\"obius function on $\Q(N)$ is given by, for any $(A,A'),(B,B')\in \Q(N)$
\[
\mu((A,A'),(B,B')) = \left\{    \begin{array}{ll}
                                (-1)^{|B\setminus A| + |A'\setminus B'|}, &
                                \text{ if } (A,A')\sqsubseteq (B,B') \text{ and }
                                A'\cap B= \emptyset\\ 
                                0, & \text{ otherwise.}
                                \end{array}     \right.
\]
\end{theorem}  
\begin{proof}
We use the fact that if $P,Q$ are posets, then the M\"obius function on
$P\times Q$ with the product order is the product of the M\"obius functions on
$P$ and $Q$ \cite{rot64}. In our case, this gives
\[
\mu_{3^n}((x_1,y_1),\ldots,(x_n,y_n)) = \prod_{i=1}^n \mu_3(x_i,y_i)
\]
where $\mu_{3^n}$ is the M\"obius function on $\Q(N)=3^n$, $\mu_3$ the M\"obius
function on $3:=\{-1,0,1\}$, and $(x_1,\ldots,x_n), (y_1,\ldots,y_n)\in
\{-1,0,1\}^n$ correspond to $(A,A'), (B,B')$ respectively. It is easy to see
that
\[
\mu_3(x_i,y_i) = \begin{cases}
                1, & \text{ if } x_i=y_i\\
                -1, & \text{ if } x_i=y_i-1\\
                0, & \text{ otherwise.}
                \end{cases}
\]
Then $\mu_{3^n}((x_1,y_1),\ldots,(x_n,y_n)) = 0 $ iff there is some $i\in N$
such that $\mu_3(x_i,y_i) =0$. This conditions reads $x_i>y_i$ or $x_i=-1,
y_i=1$. In term of subsets, this means $(A,A')\not\sqsubseteq(B,B')$ or $B\cap
A'\neq\emptyset$. 

We have $\mu_{3^n}((x_1,y_1),\ldots,(x_n,y_n)) = 1 $ iff there is no $i\in N$
such that $\mu_3(x_i,y_i) =0$, and the number of $i\in N$ such that
$\mu_3(x_i,y_i) =-1$ is even. We examine the second condition. We have:
\[
\mu_3(x_i,y_i)=-1 \Leftrightarrow \begin{cases}
                                x_i=0 & \text{ and } y_i=1\\
                                \text{or} &\\
                                x_i=-1 & \text{ and } y_i=0
                                \end{cases}
\]
which in terms of subsets, reads ($|B\setminus A|=1$ and $|A'\setminus B'|=0$)
or ($|B\setminus A|=0$ and $|A'\setminus B'|=1$). Then clearly the second
condition is equivalent to $|B\setminus A|+|A'\setminus B'|=2k$. The case
$\mu_{3^n}((x_1,y_1),\ldots,(x_n,y_n)) = -1 $ works similarly. 
\end{proof}

Consequently, the M\"obius transform of $v$ is expressed by
\begin{equation}
m(A,A') = \sum_{\substack{(B,B')\sqsubseteq(A,A')\\ B'\cap A = \emptyset}}(-1)^{|A\setminus
B|+|B'\setminus A'|}v(B,B') = \sum_{\substack{B\subseteq A\\  A'\subseteq
B'\subseteq 
A^c}}(-1)^{|A\setminus
B|+|B'\setminus A'|}v(B,B').
\label{eq:bimob}
\end{equation}
By definition of the M\"obius transform, we have
\begin{equation}
v(A,A') = \sum_{(B,B')\sqsubseteq(A,A')}m(B,B').
\label{eq:zeta}
\end{equation}
These equations are valid for any real-valued function $v$ on $\Q(N)$. If $v$
is a normalized bi-capacity, we remark that $m(\emptyset,N)=
v(\emptyset,N) = -1$, and $\sum_{(A,B)\in \Q(N)}m(A,B) = v(N,\emptyset) = 1$. 
Also, 
\begin{equation}
\label{eq:mob0}
\sum_{B\subseteq N}m(\emptyset,B)=v(\emptyset,\emptyset)=0.
\end{equation}

Proceeding as in \cite{grmaro99a}, we may write the M\"obius transform into a
matrix form, using the total order we have defined on $\Q(N)$. Denoting $v,m$
put in vector form as $v_{(n)}, m_{(n)}$, Eq. (\ref{eq:bimob}) can be rewritten
as
\[
m_{(n)} = T_{(n)}\circ v_{(n)}
\]
where $\circ$ is the usual matrix product, and $T_{(n)}$ is the matrix coding
the M\"obius transform. As in the case of classical capacities, $T_{(n)}$ has
an interesting fractal structure, as it can be seen from the case $n=2$
illustrated below.
\[
T_{(2)} = 
\begin{array}{cc}
& \;\;\;\emptyset,12\;\,\emptyset,2\;\;\,1,2\;\;\,\emptyset,1\;\;\,\emptyset,\emptyset\;\;\,1,\emptyset\;\;\,2,1\;\;\,2,\emptyset\;\,12,\emptyset\\
\begin{matrix}  \emptyset,12 \\ \emptyset,2 \\ 1,2\\
\emptyset,1 \\ \emptyset,\emptyset \\ 1,\emptyset \\ 2,1 \\ 2,\emptyset \\ 12,\emptyset
\end{matrix}
& \left[\begin{array}{rrrrrrrrr}
1 &  &    & &  & & & & \\
-1&  1 &  &    & &  & & & \\
  & -1 &  1 &    &   &   & & & \\
-1&    &    & 1  & &   &  &  & \\
1 & -1 &    & -1 &  1& &    & &  \\
  &  1 &-1  &    &-1 &  1&    &   &  \\
  &    &    & -1 &   &   &  1 & & \\
  &    &    & 1  &-1 &   & -1 &  1& \\
  &    &    &    & 1 & -1&    &-1 & 1
\end{array}\right]
\end{array}
\] 
The generating element has the form
\[
\left[\begin{array}{rrr}
1 &  &    \\
-1  &  1 & \\
  &  -1  &  1 
\end{array}\right] 
\]
and is the concatenation of two generating elements
$\left[\begin{smallmatrix} 1 & \\ -1 & 1  
\end{smallmatrix}\right]$ of the M\"obius transform
for classical capacities \cite{grmaro99a}. 

Let us examine several particular cases of bi-capacities.
\begin{proposition}
\label{prop:mcpt}
Let $v$ be a bi-capacity of the CPT type, with $v(A,B) =
\nu_1(A)-\nu_2(B)$. Then its M\"obius transform is given by:
\begin{align*}
m(A,A^c) & = m^{\nu_1}(A), \quad \forall A\subseteq N, A\neq\emptyset\\
m(\emptyset,B) & = m^{\overline{\nu_2}}(B^c), \quad \forall B\subsetneq N\\
m(\emptyset,N) & = -1\\
m(A,B) & = 0, \quad \forall (A,B)\in \Q(N) \text{ such that } A\neq\emptyset
\text{ and } B\neq A^c.
\end{align*}
\end{proposition}
\begin{proof}
Let us consider $A\neq\emptyset$. We have 
\[
m(A,A')  = \sum_{A'\subseteq B'\subseteq A^c}(-1)^{|B'\setminus
A'|}\left[\sum_{B\subseteq A}(-1)^{|A\setminus B|}v(B,B')\right].
\]
\begin{align*}
\sum_{B\subseteq A}(-1)^{|A\setminus B|}v(B,B') & = \sum_{B\subseteq
A}(-1)^{|A\setminus B|}(\nu_1(B)-\nu_2(B'))\\
        & = \sum_{B\subseteq A}(-1)^{|A\setminus B|}\nu_1(B) -
\nu_2(B')\sum_{B\subseteq A}(-1)^{|A\setminus B|}\\
        & = \sum_{B\subseteq A}(-1)^{|A\setminus B|}\nu_1(B) = m^{\nu_1}(A),
\end{align*}
where we have used (\ref{eq:sum}). Putting in $m(A,A')$ leads to 
\[
m(A,A') = m^{\nu_1}(A)\sum_{A'\subseteq B'\subseteq A^c}(-1)^{|B'\setminus
A'|}.
\]
Using again (\ref{eq:sum}), the sum is zero unless $A'=A^c$ (only one term in
the sum). Hence we get
\[
m(A,A') = \begin{cases}
                m^{\nu_1}(A), & \text{ if } A'=A^c\\
                0, & \text{otherwise}.
                \end{cases}
\]

Let us now take $A=\emptyset$. We have:
\begin{align*}
m(\emptyset,A')   & = \sum_{A'\subseteq B'\subseteq N}(-1)^{|B'\setminus
A'|}v(\emptyset,B') \\
        & = -\sum_{B'\supseteq A'}(-1)^{|B'\setminus
A'|}\nu_2(B').
\end{align*}
Let us consider $A'\neq N$, since in this case we know already that
$m(\emptyset,N) = -1$.  We recall that the co-M\"obius transform
\cite{grmaro99a} of a capacity $\nu$ is defined by
\[
\check{m}^{\nu}(A) =\sum_{B\supseteq A^c}(-1)^{|N\setminus B|}\nu(B).
\]
We remark that $m(\emptyset,A') = (-1)^{|N\setminus
A'|+1}\check{m}^{\nu_2}({A'}^c)$. Using the fact that $\check{m}^{\nu}(A) =
(-1)^{|A|+1}m^{\overline{\nu}}(A)$ for any $A\neq\emptyset$ \cite{grla00a}, we
finally get $m(\emptyset, A) = m^{\overline{\nu_2}}(A^c)$.
\end{proof}

We get as immediate corollaries the expression of the M\"obius transform of
symmetric and asymmetric bi-capacities. Observe in particular that for
asymmetric bi-capacities $v(A,B)= \nu(A)-\overline{\nu}(B)$, we have for any
$A\neq N$
\[
m(\emptyset, A) = m^{\nu}(A^c).
\] 
Applying the above result leads easily to the following one.
\begin{proposition}
Let $v$ be an additive bi-capacity on $\Q(N)$. Then its M\"obius transform is
non null only for the join-irreducible elements and the bottom of
$\Q(N)$. Specifically,
\begin{align*}
m(i,i^c ) & = \nu_1(i), \forall i\in N\\
m(\emptyset,i^c ) & = \nu_2(i), \forall i\in N\\
m(\emptyset, N) & = -1.
\end{align*} 
\end{proposition}
Let us remark that this result is in accordance with the result on (classical)
capacities, since the   join-irreducible elements for capacities are
precisely the singletons (atoms of the Boolean lattice).  
\begin{remark}
\label{rem:irr}
The above result suggests that join-irreducibles elements of the form
$(i,i^c)$ correspond to the positive part (we may call them by analogy
\emph{positive singletons}), while those of the form
$(\emptyset,i^c)$ correspond to the negative part (\emph{negative singletons}). 
\end{remark}

Having expressed the M\"obius transform of bi-capacities, we are in position to
introduce $k$-additive bi-capacities. Our definition of 1-additive
bi-capacities should coincide with additive bi-capacities, hence the following
definition seems to make sense.  
\begin{definition}
A bi-capacity is said to be \emph{$k$-additive} for some $k$ in
$\{1,\ldots,n-1\}$ if its M\"obius transform
vanishes for all elements $(A,B)$ in $\Q^{[l]}(N)$, for $l=k+1,\dots,n$.
\end{definition} 
Equivalently, $v$ is $k$-additive iff $m(A,B)=0$ whenever $|B|<n-k$.

\section{Derivatives of bi-capacities}
\label{sec:der}
Since the derivative plays a central role in the definition of interaction,
we have to define it for bi-capacities. We start as in the classical case with
pseudo-Boolean functions. 

As pseudo-Boolean are another view of capacities, we introduce \emph{ternary
pseudo-Boolean functions} in order to recover bi-capacities. These are
simply functions $f:\{-1,0,1\}^n\longrightarrow \mathbb{R}$, and the
correspondence with bi-capacities is done in the same way as for capacities,
i.e., $f(1_S,-1_T)\equiv v(S,T)$, for any $(S,T)\in \Q(N)$. 

As variables in ternary pseudo-Boolean functions take values in $\{-1,0,1\}$,
we may think of the following quantities to define the derivative w.r.t. $i$:
$f(x_1,\ldots,x_{i-1},1,x_{i+1},\ldots,x_n) -
f(x_1,\ldots,x_{i-1},0,x_{i+1},\ldots,x_n)$, and
$f(x_1,\ldots,x_{i-1},0,x_{i+1},\ldots,x_n) -
f(x_1,\ldots,x_{i-1},-1,x_{i+1},\ldots,x_n)$. Translated into functions on
$\Q(N)$, this gives respectively the following expressions:
\begin{align}
\Delta_{i,\emptyset} v(S,T) &:=v(S\cup i,T) - v(S,T), \quad \forall
(S,T)\in\Q(N\setminus i).\\
\Delta_{\emptyset,i} v(S,T) &:=v(S,T\setminus i) - v(S,T), \quad \forall
(S,T)\in\Q(N), i\in T.
\end{align}
We call them respectively \emph{left derivative} and \emph{right
derivative}. The notation and names are self-explanatory, if we remember that
in $\Q(N)$, the left (resp. right) argument concerns the positive
(resp. negative) part. 

In case of bi-capacities, the monotonicity of $v$ entails that the
derivatives are non negative.  

Left and right derivatives permit to define in a recursive way the derivative
with respect to any number of right and left elements of $N$:
\begin{equation}
\label{eq:recder}
\Delta_{S,T}v(K,L):=\Delta_{i,\emptyset}(\Delta_{S\setminus i,T}v(K,L)) =
\Delta_{\emptyset,i}(\Delta_{S,T\setminus i}v(K,L)), \forall (K,L)\in
\Q(N\setminus S), L\supseteq T. 
\end{equation}
We have for example
\begin{align*}
\Delta_{i,j}v(K,L) & = v(K\cup i,L\setminus j)-v(K\cup i,L)- v(K,L\setminus j)+v(K,L)\\
\Delta_{ij,\emptyset}v(K,L) & = v(K\cup ij,L)-v(K\cup i,L)-v(K\cup j,L) +
v(K,L). 
\end{align*}
The general expression for the $(S,T)$-derivative is given by, for any
$(S,T)\in\Q(N)$, $(S,T)\neq(\emptyset,\emptyset)$:
\begin{equation}
\Delta_{S,T}v(K,L) = \sum_{\substack{S'\subseteq S\\ T'\subseteq
T}}(-1)^{(s-s')+(t-t')}v(K\cup S',L\setminus T'),\quad\forall (K,L)\in
\Q(N\setminus S), L\supseteq T.
\end{equation}
Observe that the above expression is defined even if
$(S,T)=(\emptyset,\emptyset)$, and leads to $\Delta_{\emptyset,\emptyset}v\equiv
v$, which seems natural.
\begin{remark}
\label{rem:der}
Using Remark 1, we are tempted to consider the left derivative w.r.t. $i$ as a
derivative w.r.t the element $(i,i^c)$ of $\Q(N)$, and the right derivative as a
derivative w.r.t $(\emptyset,i^c)$. This view is supported in
\cite{grla03b,grla03c}, and serves as a basis for a general definition of
derivatives of functions on lattices. We denote them $\Delta_{(i,i^c)}$ and
$\Delta_{(\emptyset,i^c)}$ to distinguish from our previous notation. Although
less intuitive, this notation will more easily reveal structures, as we will
show later. The correspondence between the two expressions are
$\Delta_{S,T}\equiv\Delta_{(S,N\setminus(S\cup T))}$ and $\Delta_{(S,T)}\equiv
\Delta_{S,N\setminus(S\cup T)}$.
\end{remark}

We express the derivative in terms of the M\"obius transform. The starting
point is the following.
\begin{lemma}
\label{lem:der}
For any $i\in N$,
\begin{align}
\Delta_{i,\emptyset}v(S,T) = & \sum_{(S',T')\in [(i,i^c),(S\cup
i,T)]}m(S',T'), \quad \forall (S,T)\in \Q(N\setminus i) \label{eq:derl}\\
\Delta_{\emptyset, i}v(S,T) = & \sum_{(S',T')\in
[(\emptyset,i^c),(S,T\setminus i)]}m(S',T'), \quad \forall (S,T)\in \Q(N), T\ni
i\label{eq:derr}
\end{align}
\end{lemma}
\begin{proof}
Let us show (\ref{eq:derl}). For any $(S,T)\in \Q(N\setminus i)$,
\begin{align*}
\Delta_{i,\emptyset}v(S,T) = & v(S\cup i,T) - v(S,T)\\
        = & \sum_{(S',T')\sqsubseteq (S\cup i, T)}m(S',T') - \sum_{(S',T')\sqsubseteq
        (S,T)}m(S',T')\\
        = & \sum_{(S',T')\sqsubseteq (S,T)}m(S'\cup i,T').
\end{align*}
On the other hand, 
\begin{align*}
[(i,i^c),(S\cup i,T)] = & \{(S',T')\in \Q(N)|i\in S'\subseteq S\cup i,
T\subseteq 
T'\subseteq i^c\}\\
        = & \{(S'\cup i,T')\in \Q(N)|S'\subseteq S,T'\supseteq T\}
\end{align*}
hence the result. Similarly, we have
\begin{align*}
\Delta_{\emptyset, i}v(S,T) = & v(S,T\setminus i) - v(S,T)\\
        = & \sum_{(S',T')\sqsubseteq (S, T\setminus i)}m(S',T') - \sum_{(S',T')\sqsubseteq
        (S,T)}m(S',T')\\
        = &  \sum_{\substack{S'\subseteq S, T'\supseteq T, i\not\in T'\\ T'\cap
        S'=\emptyset}}m(S',T') = 
        \sum_{(S',T')\in [(\emptyset,i^c),(S,T\setminus i)]}m(S',T').
\end{align*}
\end{proof}

By induction, one can show the following general result.
\begin{proposition}
\label{prop:der}
For any $(\emptyset,\emptyset)\neq(S,T)$ in $\Q(N)$,
\begin{equation}
\Delta_{S,T}v(K,L) = \sum_{(S',T')\in[\bigsqcup\limits_{i\in S}(i,i^c)\sqcup
\bigsqcup\limits_{j\in T}(\emptyset,j^c),(S\cup K,L\setminus
T)]}m(S',T'),\quad\forall (K,L)\in \Q(N\setminus S), L\supseteq T
\label{eq:derm}
\end{equation}
\end{proposition}
\begin{proof}
We prove (\ref{eq:derm}) by induction over $(S,T)$. The result holds for
$(i,\emptyset)$ and $(\emptyset,i)$ due to Lemma \ref{lem:der}. We suppose that
the above formula holds up to a given cardinality of $S$ and $T$. Let us
compute $\Delta_{S\cup k,T}v(K,L)$, for some $k\in N\setminus(S\cup T)$, and
any $(K,L)\in \Q(N\setminus(S\cup k))$, $L\supseteq T$.  We use the fact that
(see (\ref{eq:irre}))
\[
\bigsqcup\limits_{i\in S}(i,i^c)\sqcup \bigsqcup\limits_{j\in T}(\emptyset,j^c)
= (S, N\setminus (S\cup T)).
\]
We have
\begin{align*}
\Delta_{S\cup k,T}v(K,L) & = \Delta_{k,\emptyset}(\Delta_{S,T}v(K,L)) =
\Delta_{S,T}v(K\cup k,L) - \Delta_{S,T}v(K,L)\\
        & = \sum_{(S',T')\in[(S,N\setminus (S\cup T)), (S\cup K\cup
k, L\setminus T)]}m(S',T') - \sum_{(S',T')\in[(S,N\setminus (S\cup T)), (S\cup
K, L\setminus T)]}m(S',T')\\
        & = \sum_{\substack{S\subseteq S'\subseteq S\cup K\cup k\\
L\setminus T\subseteq T'\subseteq N\setminus(S\cup T)}}m(S',T') -
\sum_{\substack{S\subseteq 
S'\subseteq S\cup K\\ L\setminus T\subseteq T'\subseteq N\setminus(S\cup
T)}}m(S',T')\\ 
& = \sum_{\substack{S\cup k\subseteq S'\subseteq S\cup K\cup k\\
L\setminus T\subseteq 
T'\subseteq N\setminus(S\cup T\cup k)\\T'\cap S'=\emptyset}}m(S',T')\\
& =  \sum_{(S',T')\in [(S\cup k,N\setminus(S\cup T\cup k)), (S\cup K\cup
k, L\setminus T)]}m(S',T')\\
&  =
 \sum_{(S',T')\in [\bigsqcup\limits_{i\in S\cup k}(i,i^c)\sqcup \bigsqcup\limits_{j\in T}(\emptyset,j^c), (S\cup K\cup
k, L\setminus T)]}m(S',T') 
\end{align*}
which is the desired result. The case of $\Delta_{S,T\cup k}v(K,L)$ works
similarly. 
\end{proof}

Remark that for any $(S,T)\in \Q(N)$, 
\[
m(S,T)   = \Delta_{(S,T)}v(\emptyset, N\setminus S).
\]
Indeed, using the above proposition
\[
\Delta_{(S,T)}v(\emptyset,N\setminus S) = \Delta_{S,N\setminus (S\cup
T)}v(\emptyset,N\setminus S) = \sum_{(S',T')\in[(S,T),(S,T)]}m(S',T') = m(S,T).
\]
This generalizes the classical result on M\"obius transform of capacities (see
Eq. (\ref{eq:md})). 

\section{Shapley value and interaction index}
\label{sec:bicoop}

\subsection{Introduction}
\label{sec:bcintro}
We consider now bi-capacities as games, i.e., the monotonicity assumption (ii)
of Def. \ref{def:bica} is no more required. We could call such games
\emph{bi-cooperative games}, as Bilbao \emph{et al.} \cite{bil00}.
Let us denote by $\mathcal{G}^{[2]}(N)$ the set of all bi-cooperative games on
$N$, and by $\mathcal{G}^{[2]}:=\bigcup_{N|n\in
\mathbb{N}^*}\mathcal{G}^{[2]}(N)$ the set of all bi-cooperative games with a
finite number of players.

An example of bi-cooperative game is the one of
\emph{ternary voting games} as proposed by Felsenthal and Machover
\cite{fema97}, where the value of $v$ is limited to $\{-1,1\}$. In ternary
games, $v(S,T)$ for any $(S,T)\in\Q(N)$ is interpreted as the result of voting
(+1: the bill is accepted, $-1$: the bill is rejected) when $S$ is the set of
voters voting in favor and $T$ the set of voters voting against. $N\setminus
S\cup T$ is the set of abstainers.

For (general) bi-cooperative games, one can keep the same kind of
interpretation: $v(S,T)$ is the worth of coalition $S$ when $T$ is the
opposite coalition, and $N\setminus S\cup T $ is the set of indifferent
(indecise) players. We call $S$ the defender coalition, and $T$ the defeater
coalition. Hence, a bi-cooperative game $v$ reduces to an ordinary cooperative
game $\nu$ if it is equivalent to know either the defender coalition $S$ or the
defeater coalition $T$, i.e. $v(S,T)=v(S,T')=:\nu(S)$ for all $T,T'\subset
N\setminus S$, or $v(S,T)= v(S',T)=:\nu(N\setminus T)$ for all $S,S'\subset
N\setminus T$.

An important concept in game theory is the Shapley value \cite{sha53} and other
related indices (e.g., Banzhaf index, probabilistic values), as well as their
generalizations as interaction indices \cite{gra96f,grmaro99a}. Our aim is to
introduce corresponding notions for bi-cooperative games, and to express them
in terms of derivatives and the M\"obius transform. Since the axiomatic
construction of the proposed notions is rather long and is itself a whole topic
(see \cite{lagr04} for the detailed axiomatic construction), we will
just cite the underlying axioms, and focus on the expressions in terms of
derivative and M\"obius transform.

We begin by recalling basic definitions and facts for (classical) cooperative
games.  For any cooperative game $\nu$ on $N$, the \emph{Shapley value} is the
vector $(\phi_1^{\nu},\ldots,\phi_n^{\nu})$, with
\[
\phi_i^{\nu} = \sum_{S\subseteq N\setminus i} \frac{(n-s-1)!s!}{n!}(\nu(S\cup i)-\nu(S)).
\] 
A single component $\phi_i^{\nu}$ is usually called the \emph{Shapley index} of
$i$.  Among remarkable properties we have that $\phi_i$ is a linear operator on
the set of cooperative games, and $\sum_{i=1}^n \phi_i^{\nu} = 1$. The
\emph{interaction index} generalizes the Shapley index, and is defined by:
\begin{equation} 
\label{eq:int}
I^{\nu}(S):=\sum_{T\subseteq N\setminus
S}\frac{(n-t-s)!t!}{(n-s+1)!}\sum_{K\subseteq 
S}(-1)^{s-k}\nu(K\cup T),\forall S\subseteq N.
\end{equation}
We have $\phi^\nu_i = I^{\nu}(\{i\})$ for all $i\in N$. The interaction index
has been axiomatized in a way similar to the Shapley value \cite{grro97a}. 

The interaction index can be expressed in a compact form using the derivative
of $\nu$ (see Section \ref{sec:prel}):
\[
I^{\nu}(S) = \sum_{T\subseteq N\setminus
S}\frac{(n-t-s)!t!}{(n-s+1)!}\Delta_S\nu(T).
\]
The expression of the interaction index in terms of the M\"obius transform of
$\nu$ is even simpler \cite{gra96f}:
\begin{equation}
\label{eq:intmob}
I^\nu(S) = \sum_{T\supseteq S}\frac{1}{t-s+1}m(T), \quad\forall S\subseteq N.
\end{equation}
This expression shows that for $k$-additive capacities, $I(S)=0$ for any
$S\subseteq N$ such that  $|S|>k$, and $I(S)=m(S)$ when $|S|=k$. Interaction for
the conjugate game is given by \cite{gra96f}:
\begin{equation}  
\label{eq:conju}
I^{\overline{\nu}}(S) = (-1)^{s+1}I^\nu(S).
\end{equation}

\subsection{Bi-unanimity games}
A direct transposition of the notion of unanimity game leads to the
following. Let $(S,S')$ in $\Q(N)$. The \emph{bi-unanimity game} centered on
$(S,S')$ is defined by:
\begin{equation}
\label{eq:biuna}
u_{(S,S')}(T,T') = \begin{cases}
                1, & \text{ if } T\supseteq S \text{ and } T'\subseteq S'\\
                0, & \text{ otherwise}.
                \end{cases}
\end{equation}
Hence, as in the classical case, the set of all bi-unanimity games is a basis
 for bi-capacities:
\begin{equation}
\label{eq:una}
v(T,T') = \sum_{(S,S')\in\Q(N)}m(S,S')u_{(S,S')}(T,T').
\end{equation}
Remark that $u_{(S,S')}$ is not a normalized bi-capacity since
 $u_{(S,S')}(\emptyset,N)\neq -1$, and $u_{(\emptyset,N)}$ is not a bi-capacity
 since $u_{(\emptyset,N)}(\emptyset,\emptyset)=1$.

It is easy to see by (\ref{eq:zeta}) that the M\"obius transform of
$u_{(S,S')}$ is
\[
m^{u_{(S,S')}}(T,T') = \begin{cases}
                        1, & \text{ if } (T,T') = (S,S')\\
                        0, & \text{otherwise}.
                        \end{cases}
\]

\subsection{The Shapley value for bi-cooperative games}
\label{sec:shabi}
In classical games, the Shapley value expresses the contribution of each player
in the game, or more precisely the average difference between the situations
where some player $i$ participates to the game or does not participate. In the
case of bi-cooperative games, since each player can join either the defender or
the defeater part, besides no participation, we should define a Shapley value
for the case when players join the defender part, and another one when players
join the defeater part, instead of a single value. We denote by
$\phi^v_{i,\emptyset}$ and $\phi^v_{\emptyset,i}$ the coordinates of the
Shapley value for player $i$ for the defender and defeater part
respectively. Hence, we consider the Shapley value as an operator on the set of
bi-cooperative games $\phi:\mathcal{G}^{[2]}(N)\longrightarrow
\mathbb{R}^{2n}$~; $v\mapsto \phi^v$, for any finite support $N$, and
coordinates of $\phi^v$ are either of the $\phi^v_{i,\emptyset}$ or
$\phi^v_{\emptyset,i}$ type.

We present briefly the axioms giving rise to our definition, without details
(see \cite{lagr04}).  
\begin{quote}
\textbf{Linear axiom (L):} $\phi^v$ is linear on the set of games
$\mathcal{G}^{[2]}(N)$. 
\end{quote}
Player $i$ is said to be \emph{left-null} (resp. \emph{right-null}) if $v(S\cup
i,T)=v(S,T)$ (resp. $v(S,T\cup i)=v(S,T)$) for all $(S,T)\in \Q(N\setminus i)$.
\begin{quote}
  \textbf{Left-null axiom (LN):} $\forall v\in\mathcal{G}(N)$, for all $i\in
  N$, $\phi^v_{i,\emptyset}=0$ if $i$ is left-null.
\end{quote}
\begin{quote}
  \textbf{Right-null axiom (RN):} $\forall v\in\mathcal{G}(N)$, for all $i\in
  N$, $\phi^v_{\emptyset,i}=0$ if $i$ is right-null.
\end{quote}
Let $\sigma$ be a permutation on $N$. With some abuse of notation, we denote
$\sigma(S):=\{\sigma(i)\}_{i\in S}$.
\begin{quote}
\textbf{Fairness axiom (F):} $\phi^{v\circ\sigma^{-1}}_{\sigma(i),\emptyset}
= \phi^v_{i,\emptyset}$, and $\phi^{v\circ\sigma^{-1}}_{\emptyset,\sigma(i)} =
\phi^v_{\emptyset,i}$, for all $i\in N$, for all $v\in \mathcal{G}(N)$.
\end{quote}
This axiom, usually called ``symmetry axiom'', says that $\phi^v$ should not
depend on the labelling of the players.
\begin{quote}
\textbf{Symmetry axiom (S):} Let us consider $v_1,v_2$ in $\mathcal{G}(N)$ such
that the following holds for some $i\in N$:
\[
v_2(S\cup i,T)-v_2(S,T) = v_1(S,T) - v_1(S,T\cup i),
\quad\forall(S,T)\in\Q(N\setminus i).
\] 
Then $\phi^{v_2}_{i,\emptyset} = \phi^{v_1}_{\emptyset,i}$. 
\end{quote}
The axiom says that when a game $v_2$ behaves symmetrically with $v_1$ (in
the sense of inverting left and right arguments, up to the sign), then the
Shapley values are the same. It means that the ways the computation is done for
left and right parts are identical.
\begin{quote}
\textbf{Efficiency axiom (E):} $\sum_{i\in N}\big(\phi^v_{i,\emptyset} +
\phi^v_{\emptyset,i}\big) = v(N,\emptyset) - v(\emptyset, N)$.
\end{quote}
\begin{quote}
\textbf{Unanimity game axiom (UG):} for any unanimity game $u_{(S,T)}$,
\begin{align*}
\phi^{u_{(S,T)}}_{i,\emptyset} & =\begin{cases}
              \frac{1}{n-t},& \text{ if } i\in S\\
              0, & \text{ otherwise}
              \end{cases}\\
\phi^{u_{(S,T)}}_{\emptyset,i} & =\begin{cases}
              \frac{1}{n-t},& \text{ if } i\in N\setminus(S\cup T)\\
              0, & \text{ otherwise}.
              \end{cases}\\
\end{align*}
\end{quote}
Axiom \textbf{(UG)} says that all players in $S$ are equally important and
others are not important for the defender part (since $S$ is the set of
necessary players in the defender part for winning), while for the defeater
part, only players outside $S$ and $T$ are important (since they may make the
game equal to 0 if they become defeaters).  Now, the total value of the game is
to be shared among all players except those of $T$ since they are not decisive,
hence the amount given to each player is $\frac{1}{n-t}$.
\begin{theorem}
\cite{lagr04}\\
(i)  Under axioms \textbf{(L)}, \textbf{(LN)}, \textbf{(RN)}, \textbf{(F)},
  \textbf{(S)} and \textbf{(E)}, 
\begin{align*}
\phi^v_{i,\emptyset} & = \sum_{S\subseteq N\setminus i}
\frac{(n-s-1)!s!}{n!}[v(S\cup 
i,N\setminus(S\cup i)) - v(S,N\setminus(S\cup i))]\\
\phi^v_{\emptyset,i} & = \sum_{S\subseteq N\setminus i}
\frac{(n-s-1)!s!}{n!}[v(S,N\setminus(S\cup i)) - v(S,N\setminus S)].
\end{align*}
(ii) Under axioms  \textbf{(L)}, \textbf{(LN)}, \textbf{(RN)}, and
\textbf{(F)}, axioms \textbf{(S)} and \textbf{(E)} are equivalent to
\textbf{(UG)}. 
\end{theorem}
Using derivatives, a more compact form is
\begin{align}
\phi^v_{i,\emptyset} & = \sum_{S\subseteq N\setminus i} \frac{(n-s-1)!s!}{n!}\Delta_{i,\emptyset}v(S,N\setminus(S\cup i))\label{eq:sha+}\\
\phi^v_{\emptyset,i} & = \sum_{S\subseteq N\setminus i}
\frac{(n-s-1)!s!}{n!}\Delta_{\emptyset,i}v(S,N\setminus S)\label{eq:sha-}.
\end{align} 

It is easy to see that if $v$ is of the CPT type, i.e. $v(S,T) = \nu_1(S)
- \nu_2(T)$, then
\begin{align*}
\phi^v_{i,\emptyset} & = \phi^{\nu_1}(i)\\
\phi^v_{\emptyset,i} & =\phi^{\nu_2}(i),
\end{align*}
where $\phi^{\nu_1},\phi^{\nu_2}$ are the (classical) Shapley values of $\nu_1$
and $\nu_2$.

The following expression gives the Shapley value in terms of the M\"obius
transform. 
\begin{proposition}
\label{prop:sham}
Let $v$ be a bi-cooperative game on $N$. For any $i\in N$,
\begin{align*}
\phi^v_{i,\emptyset} &= \sum_{(S,T)\sqsupseteq(i,i^c)}\frac{1}{n-t}m(S,T)\\
\phi^v_{\emptyset,i}
&=\sum_{(S,T)\sqsupseteq(\emptyset,i^c),(S,T)\in\Q(N\setminus
  i)}\frac{1}{n-t}m(S,T). 
\end{align*}
\end{proposition}
This result will be a particular case of a more general result (see Prop. \ref{prop:biint}).

\subsection{The interaction index}
\label{sec:intbi}
For classical games, the interaction index $I^\nu(S)$ can be obtained from the
Shapley value $\phi^\nu(i)=:I^\nu(\{i\})$ by a recursion formula
\cite{grro97a}. We take here a similar approach, and propose recursion formulas
which are exact counterparts of the one for classical games. They will permit
to build $I^v_{S,\emptyset}$ and $I^v_{\emptyset, T}$ from
$\phi^v_{i,\emptyset}=:I^v_{i,\emptyset}$ and
$\phi^v_{\emptyset,i}=:I^v_{\emptyset,i}$ respectively. However, to build
$I^v_{S,T}$ for any $(S,T)\in \Q(N)$, we need a third starting point which is
$I^v_{i,j}$, yet to be defined. In this paper, we define it by analogy with
interaction for classical games, an axiomatic approach being out of our scope
here. This approach is detailed in \cite{grla03c}.

Taking the elementary case where $n=2$, observe that the interaction index for
some classical game $\nu$ reduces to (see (\ref{eq:int})):
\[
I^\nu(\{1,2\}) = \nu(\{1,2\}) - \nu(\{1\}) - \nu(\{2\}) + \nu(\emptyset).
\]
Recalling that $\nu(A)$ is the score of the binary alternative
$(1_{A},0_{A^c})$, we see that the above expression is the difference between
alternatives on the diagonal (i.e. (1,1), the best one, and (0,0), the worst
one) and on the anti-diagonal (i.e. (1,0) and (0,1)). We keep the same scheme
and define for a bi-cooperative game $v$
\[
I^v_{1,2}:= v(\{1\},\emptyset) - v(\emptyset,\emptyset) - v(\{1\},\{2\}) + v(\emptyset,\{2\})
\]
which is in fact $\Delta_{1,2}v(\emptyset,\{2\})$. Hence we are lead naturally
to the following, in the general case:
\begin{equation}
\label{eq:ij}
I^v_{i,j} = \sum_{S\subseteq
  N\setminus i,j}\frac{(n-s-2)!s!}{(n-1)!}\Delta_{i,j}v(S,N\setminus (S\cup i)).
\end{equation}
We introduce necessary notions for the recursion formulas. Let $v$ be a
bi-cooperative game on $N$, and let $\emptyset\neq K\subseteq N$.  The
\emph{reduced game} $v_{[K]}$ is the game where all players in $K$ are
considered as a single player denoted $[K]$, i.e., the set of players is then
$N_{[K]}:=(N\setminus K)\cup \{[K]\}$. The reduced game is defined by
\[
v_{[K]}(S,T):=v(\eta_{[K]}(S), \eta_{[K]}(T))
\]
for any $(S,T)\in\Q(N_{[K]})$, and $\eta_{[K]}:N_{[K]}\longrightarrow N$ is
defined by
\[
\eta_{[K]}(S) :=\begin{cases}
                S, & \text{ if } [K]\not\in S\\
                (S\setminus [K])\cup K, & \text{ otherwise.}
                \end{cases}
\]
We introduce two \emph{restricted games} $v^{N\setminus K}_0$ and $v^{N\setminus
K}_-$, which are defined on $N\setminus K$, and which are linked to $v$ as
follows, for all $(S,T)\in\Q(N\setminus K)$:
\begin{align*}
v^{N\setminus K}_0(S,T) & := v(S,T)\\ 
v^{N\setminus K}_-(S,T) & := v(S,T\cup K).
\end{align*}
$v^{N\setminus K}_0$ is a restriction of $v$ where all players in $K$ are
neutral (0 level), while for $v^{N\setminus K}_-$, all players in $K$ are
defeaters (hence the ``$-$'' sign).

The interaction index is an operator $I$ on the set of games
$\mathcal{G}^{[2]}(N)\longrightarrow \mathbb{R}^{\Q(N)}$ ; $v\mapsto I^v$, for
any finite support $N$. We denote by $I^v_{S,T}$ the interaction index when $S$
is added to the defender coalition, and $T$ is withdrawn from the defeater
coalition.

The following recursion formulas are direct transpositions of what was proposed
for classical games in \cite{grro97a}.
\begin{quote}
\textbf{Recursivity (R):} for any $v\in\mathcal{G}^{[2]}$,
\begin{align}
I^v_{S,T} & = I^{v_{[S]}}_{[S],T} - \sum_{K\subsetneq
S,K\neq\emptyset}I^{v^{N\setminus K}_0}_{S\setminus K,T},\quad \forall
(S,T)\in \Q(N),S\neq\emptyset\label{eq:rec+}\\ 
I^v_{S,T} & = I^{v_{[T]}}_{S,[T]} - \sum_{K\subsetneq
T,K\neq\emptyset}I^{v^{N\setminus K}_-}_{S,T\setminus K},\quad \forall
(S,T)\in \Q(N),T\neq\emptyset\label{eq:rec-} 
\end{align}
\end{quote} 
Applying these formulas, we get the following expression for the interaction
index.
\begin{theorem}
  Suppose that the interaction index $I^v_{S,T}$ is such that
  $I^v_{i,\emptyset}$, $I^v_{\emptyset,i}$ and $I^v_{i,j}$ are given by
(\ref{eq:sha+}), (\ref{eq:sha-}) and (\ref{eq:ij}). Then $I^v$ satisfies
\textbf{(R)} if and only if
\begin{equation}
\label{eq:biint}
I^v_{S,T}=\sum_{K\subseteq N\setminus (S\cup T)}\frac{(n-s-t-k)!k!}{(n-s-t+1)!}
\Delta_{S,T}v(K,N\setminus(K\cup S)),
\end{equation}
for all $(S,T)\in\Q(N)$, $(S,T)\neq(\emptyset,\emptyset)$.
\end{theorem}
\begin{proof}
The if part is left to the reader. To prove the converse, we proceed by a
double induction on $|S|$ and $|T|$. Clearly, the formula is true for
$I^v_{i,\emptyset}$, $I^v_{\emptyset,i}$, and $I^v_{i,j}$. Let us assume it is
true up to $|S|=s-1$ and $|T|=t-1$. We will prove that if $s\geq 1$, it is still
true for $s$ and $t-1$, and if $t\geq 1$, it is still true for $s-1$ and
$t$. This suffices to show the result for any $(S,T)\in\Q(N)$,
$(S,T)\neq(\emptyset,\emptyset)$. 

By induction assumption we have, for any $S\subseteq N$, $|S|=s$, using
(\ref{eq:recder}):
\begin{eqnarray*}
I^{v_{[S]}}_{[S],T} & = &\sum_{K\subseteq N\setminus
  (S\cup T)}\frac{(n-s-t-k)!k!}{(n-s-t+1)!}\big[\Delta_{\emptyset,T}v(K\cup
  S,N\setminus(K\cup S)) \\ &&- \Delta_{\emptyset,T}v(K,N\setminus(K\cup S))
  \big]\\ 
I^{v^{N\setminus J}_0}_{S\setminus J,T} & = &\sum_{K\subseteq N\setminus
  (S\cup T)}\frac{(n-s-t-k)!k!}{(n-s-t+1)!}\sum_{S''\subseteq S\setminus
  J}(-1)^{s-s''-j}\Delta_{\emptyset,T}v(K\cup S'',N\setminus (K\cup S))
\end{eqnarray*} 
for any  $J\subsetneq S$, $J\neq\emptyset$. Using (\ref{eq:rec+}), we get:
\begin{multline}
I^v_{S,T} = \sum_{K\subseteq N\setminus
  (S\cup T)}\frac{(n-s-t-k)!k!}{(n-s-t+1)!}\Big[\Delta_{\emptyset,T}v(K\cup S,N\setminus(K\cup S)) -
  \Delta_{\emptyset,T}v(K,N\setminus(K\cup S))  \\
-\sum_{J\subsetneq S,J\neq\emptyset}\sum_{S''\subseteq S\setminus
  J}(-1)^{s-s''-j}\Delta_{\emptyset,T}v(K\cup S'', N\setminus (K\cup S))\Big].
\end{multline}
The last term into brackets can be rewritten as:
\begin{eqnarray*}
\lefteqn{\sum_{S''\subsetneq S}\Delta_{\emptyset,T}v(K\cup S'',N\setminus (K\cup S))\sum_{\substack{J\subseteq
  S\setminus S''\\ J\neq\emptyset,S}}(-1)^{s-s''-j}}\\
   & = & \sum_{\substack{S''\subsetneq S\\S''\neq \emptyset}}\Delta_{\emptyset,T}v(K\cup
  S'',N\setminus (K\cup S))\sum_{\substack{J\subseteq S\setminus S''\\
  J\neq\emptyset}} (-1)^{s-s''-j} + \Delta_{\emptyset,T}v(K,N\setminus (K\cup
  S))\sum_{\substack{J\subsetneq S\\ J\neq\emptyset}} (-1)^{s-j}\\
  & = & -\sum_{\substack{S''\subsetneq S\\S''\neq \emptyset}}(-1)^{s-s''}\Delta_{\emptyset,T}v(K\cup
  S'',N\setminus (K\cup S)) - (-1)^s\Delta_{\emptyset,T}v(K,N\setminus (K\cup
  S))\\ 
  & & - \Delta_{\emptyset,T}v(K,N\setminus(K\cup S)),
\end{eqnarray*}
where we have used the fact that $\sum_{S\subseteq N}(-1)^{n-s}=0$. Putting this
into the bracketted term, it becomes $\Delta_{S,T}v(K,N\setminus(K\cup
S))$, which proves the result.

The proof with $T$ works similalry. Starting expressions are, for $|T|=t$:
\begin{eqnarray*}
I^{v_{[T]}}_{S,[T]} & = &\sum_{K\subseteq N\setminus
  (S\cup
  T)}\frac{(n-s-t-k)!k!}{(n-s-t+1)!}\big[\Delta_{S,\emptyset}v(K,N\setminus(K\cup
  S\cup T)) \\
 &&- \Delta_{S,\emptyset,}v(K,N\setminus(K\cup S))
  \big]\\ 
I^{v^{N\setminus J}_-}_{S,T\setminus J} & = &\sum_{K\subseteq N\setminus
  (S\cup T)}\frac{(n-s-t-k)!k!}{(n-s-t+1)!}\sum_{T''\subseteq T\setminus
  J}(-1)^{t-t''-j}\Delta_{S,\emptyset}v(K,N\setminus (K\cup S\cup T''))
\end{eqnarray*} 
for any  $J\subsetneq T$, $J\neq\emptyset$.
\end{proof}

Since $\Delta_{\emptyset,\emptyset}v$ is defined, we extend the definition of
$I^v_{S,T}$ to the case where $(S,T)=(\emptyset,\emptyset)$.

\begin{remark}
Using Remark 1 again as we did for derivatives, we may denote the pair $S,T$ of
defender and defeater parts as the corresponding element $(S,N\setminus (S\cup
T))$ of $\Q(N)$, and thus denoting $I^v_{S,T}$ by $I^v(S,N\setminus (S\cup
T))$. Then, $I^v(S,T)$ is interpreted as the interaction when $(S,T)$ is
``added'' to some coalition $(K,L)$ by taking the supremum $(S,T)\sqcup (K,L) =
(S\cup K,T\cap L)$. Although less intuitive, let us remark that this notation,
together with the notation for derivatives introduced in Remark 2, permits to
get a much simpler expression of the bi-interaction:
\begin{equation}
\label{eq:intst}
I^v(S,T) = \sum_{K\subset T}\frac{(t-k)!k!}{(t+1)!}\Delta_{(S,T)}v(K,N\setminus(K\cup
S)).
\end{equation}
This is not surprising, since this is more in accordance with the structure of
$Q(N)$. We will sometimes use this notation, whenever it will be convenient.
\end{remark}
The expression of the interaction in terms of the M\"obius transform is given
as follows.
\begin{proposition}
\label{prop:biint}
Let $v$ be a bi-cooperative game on $N$. For any $(S,T)\subseteq \Q(N)$, 
\begin{align*}
I^v_{S,T} = & \sum_{(S',T')\in\uparrow(\bigsqcup\limits_{i\in S}(i,i^c)\sqcup
\bigsqcup\limits_{j\in T}(\emptyset,j^c))\cap \Q(N\setminus
T)}\frac{1}{n-s-t-t'+1}m(S',T')\\
 = & \sum_{(S',T')\in[(S,N\setminus(S\cup T)),(N\setminus T, \emptyset)]}\frac{1}{n-s-t-t'+1}m(S',T').
\end{align*}
\end{proposition}
To prove this result, the following combinatorial result is useful. 
\begin{lemma}
\label{lem:comb}
\[
\sum_{i=0}^k \frac{(n-i-1)!k!}{n!(k-i)!} = \frac{1}{n-k}.
\]
\end{lemma}
\begin{proof}
\begin{gather*}
\sum_{i=0}^k \frac{(n-i-1)!k!}{n!(k-i)!} =  \frac{1}{n}+\frac{k}{n(n-1)}
+\cdots+\frac{k!}{n(n-1)\cdots(n-k)}\\
=  \frac{(n-1)\cdots(n-k) + k(n-2)\cdots(n-k) + k(k-1)(n-3)\cdots(n-k)+\cdots +
k!} {n(n-1)\cdots(n-k)}.
\end{gather*}
It suffices to show that the numerator is $n(n-1)\cdots(n-k+1)$. Summing the
last two terms of the numerator, then the last three terms and so on, we get
successively:
\begin{align*}
k(k-1)\cdots 2(n-k) + k! & = k\cdots 2(n-k+1)\\
k\cdots3(n-k+1)(n-k) + k\cdots 2(n-k+1) & = k\cdots 3(n-k+1)(n-k+2)\\
\vdots  & \\
k\cdots i(n-k+i-2)\cdots(n-k) + & \\
k\cdots(i-1)(n-k+i-2)\cdots(n-k+1) & = 
                k\cdots i(n-k+i-2)\cdots\\
                & \qquad \cdots(n-k+1)(n-k+i-1)\\
\vdots  & \\
k(n-2)\cdots(n-k) + k(k-1)(n-2)\cdots(n-k+1) & = k(n-2)\cdots(n-k+1)(n-1)\\
(n-1)\cdots(n-k) + k(n-1)\cdots(n-k+1) & = (n-1)\cdots(n-k+1)n. 
\end{align*}
\end{proof}

We now prove Prop. \ref{prop:biint}. 

\begin{proof}
By Prop. \ref{prop:der}, we have
\[
\Delta_{S,T}v(K,N\setminus(K\cup S)) =  \sum_{(S'T')\in\underbrace{[(S,N\setminus(S\cup T)),(S\cup K, N\setminus (K\cup S\cup
T))]}_{\substack{S\subseteq S'\subseteq S\cup K \\ N\setminus(K\cup S\cup
T)\subseteq 
T'\subseteq N\setminus (S\cup T) \\ S'\cap T'=\emptyset}}}m(S',T').
\]
When $K=N\setminus(S\cup T)$, the interval becomes $[(S, N\setminus(S\cup
T)),(N\setminus T, \emptyset)]$, or equivalently $\uparrow(\bigsqcup\limits_{i\in
S}(i,i^c)\sqcup \bigsqcup\limits_{j\in T}(\emptyset,j^c))\cap \Q(N\setminus
T)$. This interval contains all intervals $[(S,N\setminus(S\cup T)),(S\cup K,
N\setminus (K\cup S\cup T))]$ since $S\cup K\subseteq N\setminus T$. Hence,
\begin{multline*}
\sum_{K\subseteq  N\setminus(S\cup T)}\frac{(n-s-t-k)!k!}{(n-s-t+1)!}
\Delta_{S,T}v(K, N\setminus (K\cup S)) \\
 = \sum_{(S',T')\in [(S, N\setminus(S\cup T)),(N\setminus T, \emptyset)]}
m(S',T')\sum_{\substack{K\subseteq N\setminus(S\cup T) \\ S\cup K \supseteq S' \\
N\setminus (K\cup S\cup T)\subseteq T'}}\frac{(n-s-t-k)!k!}{(n-s-t+1)!}.
\end{multline*}
Observe that in the second summation, condition $S\cup K \supseteq S'$ is
redundant. Indeed, we have $N\setminus(K \cup S\cup T)\subseteq T'
\Leftrightarrow K\cup S\cup T\supseteq N\setminus T'$.  
Since $N\setminus T'\supseteq S'$ and $T\cap S'=\emptyset$, we deduce $S\cup
K\supseteq S'$. 

Using this fact and letting $K':=N\setminus (K\cup S\cup T)$, the second
summation becomes:
\begin{align*}
\sum_{N\setminus (K\cup S\cup T)\subseteq T'}\frac{(n-s-t-k)!k!}{(n-s-t+1)!} & = 
\sum_{K'\subseteq T'}\frac{k'!(n-k'-s-t)!}{(n-s-t+1)!}\\
        & = \sum_{k'=0}^{t'}\binom{t'}{k'}\frac{k'!(n-k'-s-t)!}{(n-s-t+1)!}\\
        & = \sum_{k'=0}^{t'}\frac{t'!(n-k'-s-t)!}{(n-s-t+1)!(t'-k')!}\\
        & = \frac{1}{n-s-t-t'+1}
\end{align*}
using Lemma \ref{lem:comb}. 
\end{proof}

We examine the case of $k$-additive bi-capacities and CPT-type bi-capacities,
using at some places the notation $I^v(S,T)$ whenever this is clearer. 
\begin{proposition}
\begin{itemize}
\item [(i)]
If $v$ is a $k$-additive bi-capacity, then
\begin{align}
I^v(S,T) = & 0, \quad \forall (S,T)\in\Q(N) \text{ such that } |T|<n-k\\
I^v(S,T) = &  m(S,T), \quad \forall (S,T)\in\Q(N)
\text{ such that } |T|=n-k.
\end{align}
\item [(ii)] If $v$ is of CPT type, then $I^v_{S,T}=0$ unless $S=\emptyset$ or
$T=\emptyset$.
\item [(iii)] If $v$ is of the CPT type with $v(S,T):=\nu_1(S) - \nu_2(T)$,
then 
\begin{align*}
I^v_{S,\emptyset} &=I^{\nu_1}(S), \quad \forall S\subseteq N, S\neq\emptyset\\
I^v_{\emptyset,T} &=I^{\overline{\nu_2}}(T), \quad \forall T\subseteq N
\end{align*}
where $I^{\nu_i}$ is the classical interaction index of game $\nu_i$.
\item [(iv)] If $v$ is an asymmetric bi-capacity with underlying capacity
$\nu$, then
\[
I^v_{S,\emptyset} = I^{\nu}(S), \quad I^v_{\emptyset,T}=I^{\nu}(T)
\] 
\item [(v)] If $v$ is a symmetric bi-capacity with underlying capacity
$\nu$, then
\[
I^v_{S,\emptyset} = I^{\nu}(S), \quad I^v_{\emptyset,T}=(-1)^{t+1}I^{\nu}(T)
\] 
\end{itemize} 
\end{proposition}
\begin{proof}
(i) $v$ is $k$-additive iff $m(S',T')=0$ for all $T'$ such that $t'<n-k$. Using
Prop. \ref{prop:biint} for $I^v_{S,N\setminus(S\cup T)}$, we see that in the
summation, $T'\subseteq T$.  Consequently, if $|T|<n-k$, $m(S',T')$ will be
always 0, and so $I^v_{S,T}=0$.

Now, if $|T|=n-k$, only $T'=T$ gives a non zero term. For any $T'$, we have
$S'\subseteq N\setminus T'$. Since we have also the condition $S\subseteq
S'\subseteq S\cup T$, the only solution is $S'=S$, hence the result.

(ii) By Prop. \ref{prop:mcpt}, we know that $m(S',T')\neq 0$ iff $S'=\emptyset$
or $S'=N\setminus T'$. In the expression of $I^v_{S,T}$ of
Prop. \ref{prop:biint}, the first condition implies $S=\emptyset$, while the
second implies $T=\emptyset$. 

(iii) By Prop. \ref{prop:mcpt}, we have for any non empty subset $S$:
\begin{align*}
I_{S,\emptyset} & = \sum_{(S',T')\in[(S,N\setminus S), (N,
\emptyset)]}\frac{1}{n-s-t'+1} m(S',T')\\
        &= \sum_{\substack{S'\supseteq S, T'\subseteq N\setminus S\\
S'\cap T'=\emptyset}} \frac{1}{n-s-t'+1} m(S',T')\\
        &= \sum_{S'\supseteq S}\frac{1}{s'-s+1}m^{\nu_1}(S) = I^{\nu_1}(S)
\end{align*}
where in the last line we have used Prop. \ref{prop:mcpt}, and
(\ref{eq:intmob}). Similarly, for any subset $T$:
\begin{align*}
I_{\emptyset, T} &= \sum_{(S',T')\in[(\emptyset,N\setminus T),(N\setminus
T,\emptyset)]}\frac{1}{n-t-t'+1}m(S',T')\\
        &= \sum_{T'\subseteq N\setminus
T}\frac{1}{n-t-t'+1}m^{\overline{\nu}_2}(N\setminus T')\\
        &= \sum_{T''\supseteq T}\frac{1}{t''-t+1}m^{\overline{\nu}_2}(T'')
= I^{\overline{\nu_2}}(T).
\end{align*}

(iv) and (v) are particular cases of (iii) (use (\ref{eq:conju})).
\end{proof}

For (i), thanks to the notation $I^v(S,T)$, the comparison with the
corresponding result for capacities (see Section \ref{sec:bcintro}) is
striking.  Again if we use this notation for (iii), we obtain a result formally
identical to Prop. \ref{prop:mcpt}, replacing $I$ by $m$. This shows the
mathematical interest of this notation.

\bibliographystyle{plain}
\bibliography{../BIB/fuzzy,../BIB/grabisch,../BIB/general}

\end{document}